\documentclass[lettersize,journal,9pt]{IEEEtran}
\usepackage{amsmath,amsfonts}
\usepackage{algorithmic}
\usepackage{caption}
\usepackage{algorithm}
\usepackage{array}
\usepackage{orcidlink}
\usepackage[subrefformat=parens,labelformat=parens,font=small]{subcaption}
\usepackage{textcomp}
\usepackage{stfloats}
\usepackage{url}
\usepackage{verbatim}
\usepackage{graphicx}
\usepackage{booktabs}
\usepackage{bm}
\usepackage{cite}

\begin{document}
\title{Accelerating LLM Inference via Dynamic KV Cache Placement in Heterogeneous Memory System}

\author{Yunhua Fang\orcidlink{0009-0009-4718-8825}, Rui Xie\orcidlink{0000-0003-3177-5071}, 
        Asad Ul Haq\orcidlink{0009-0003-7975-0102}, Linsen Ma\orcidlink{0009-0000-8535-7911}, 
        Kaoutar El Maghraoui\orcidlink{0000-0002-1967-8749},\\ Naigang Wang\orcidlink{0000-0001-7664-0061}, 
        Meng Wang\orcidlink{0000-0003-0928-9691}, Liu Liu\orcidlink{0000-0003-0792-8146}, Tong Zhang\orcidlink{0009-0009-8005-0043}

\thanks{Yunhua Fang, Rui Xie, Asad Ul Haq, Linsen Ma, Meng Wang, Liu Liu, and Tong Zhang are with the Rensselaer Polytechnic Institute, Troy, NY 12180 USA.}
\thanks{Kaoutar El Maghraoui and Naigang Wang are with IBM T.J. Watson Research Center, Yorktown Heights, NY 10598 USA.}
}



\maketitle
\thispagestyle{empty}
\pagestyle{empty}
\IEEEpubidadjcol
\begin{abstract}
Large Language Model (LLM) inference is increasingly constrained by memory bandwidth, with frequent access to the key-value (KV) cache dominating data movement. While attention sparsity reduces some memory traffic, the relevance of past tokens varies over time, requiring the full KV cache to remain accessible and sustaining pressure on both bandwidth and capacity. With advances in interconnects such as NVLink and LPDDR5X, modern AI hardware now integrates high-bandwidth memory (HBM) with high-speed off-package DRAM, making heterogeneous memory systems a practical solution. This work investigates dynamic KV cache placement across such systems to maximize aggregated bandwidth utilization under capacity constraints. Rather than proposing a specific scheduling policy, we formulate the placement problem mathematically and derive a theoretical upper bound, revealing substantial headroom for runtime optimization. To our knowledge, this is the first formal treatment of dynamic KV cache scheduling in heterogeneous memory systems for LLM inference.

\end{abstract}


\section{Introduction}

\IEEEPARstart{T}{ransformer}-based large language models~\cite{Vaswani2017Attention} (LLMs) inference dominates the resource utilization of modern AI infrastructure, yet its scaling is hindered by memory bandwidth bottlenecks. These arise in the read-intensive decode stage, where KV cache accesses predominate. However, the bandwidth of memory subsystems has not kept pace with the accelerating compute capabilities of modern processors~\cite{10477550}, resulting in a mismatch that limits the inference throughput. As a result, memory bandwidth has become a fundamental constraint on the performance and scalability of LLM inference workloads. Recent research has demonstrated that enforcing attention sparsity~\cite{NEURIPS2023_6ceefa7b} is an effective way to alleviate the memory bandwidth bottleneck during LLM inference. However, because token importance evolves dynamically during decoding, the system must retain the entire KV cache in memory to ensure any token can be accessed when its relevance resurfaces~\cite{pmlr-v235-tang24l}. This demand places pressure not only on bandwidth but also on memory capacity. 

High Bandwidth Memory~(HBM) offers the highest throughput among existing memory technologies and is well-suited to address bandwidth challenges. However, its limited capacity, constrained by die stacking and packaging density, as well as its high cost and power consumption, make it undesirable as the sole memory resource for large-scale LLM inference. Consequently, heterogeneous memory systems are becoming inevitable in future AI infrastructure. Thanks to continuous advances in DRAM interfaces~(e.g., LPDDR5X) and high-speed serial interconnects~(e.g., NVLink, UALink, and PCIe), AI processors such as GPUs and TPUs can readily complement HBM with off-package DRAM fabrics that provide lower, yet increasingly comparable, bandwidth. For example, the design of NVIDIA Grace Hopper Superchip~\cite{NVIDIAGraceHopper} shows that off-package DRAM can provide bandwidth within an order of magnitude of HBM, while offering significantly greater capacity. 

\begin{figure}[!t]
  \centering
  \includegraphics[
    width=\columnwidth,
  ]{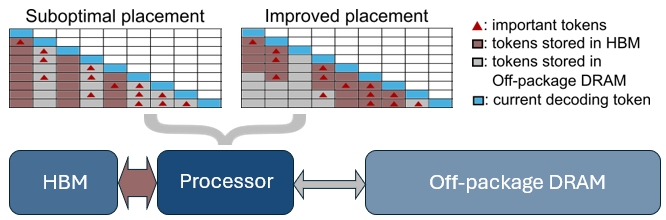}
  \caption{Illustration of suboptimal (left) vs.~improved (right) token placement across HBM and off-package DRAM. The row of the table represents the auto-regressive decoding steps, while the column represents the increasing token index.}
  \label{placement}
\end{figure}

In AI infrastructures equipped with heterogeneous memory systems, where off-package DRAM offers lower, yet comparable, bandwidth to HBM, runtime data placement scheduling becomes a critical performance factor due to the dynamic variation in token importance during LLM inference. Since LLM inference is heavily bandwidth-bound, overall throughput is closely tied to the efficient utilization of the aggregated bandwidth across the memory hierarchy. In such systems, the placement of frequently accessed tokens significantly affects performance: allocating important tokens to slower memory leads to under-utilization of high-bandwidth resources and degraded inference speed. This challenge underscores the importance of dynamic data placement strategies that adapt to evolving attention importance, as illustrated in Fig.~\ref{placement}. Yet, despite its growing relevance, data placement scheduling in heterogeneous memory systems remains largely unexplored in the context of LLM inference.

To address this gap, this work introduces a formal framework for modeling the data placement problem in heterogeneous memory systems during LLM inference. Instead of proposing a specific scheduling algorithm, we aim to characterize the design space and performance potential by deriving a theoretical upper bound through heuristic optimization. Using the classical simulated annealing (SA) algorithm, we construct a dynamic placement strategy that approximates the optimal allocation of KV cache entries based on future access patterns. Our simulation results show that this upper bound achieves up to 5.87$\times$ higher throughput compared to a static placement scheme, revealing significant headroom for improvement. By quantifying the performance gap between this bound and existing heuristics, we hope to catalyze future research on adaptive data placement scheduling strategies to unlock the full potential of heterogeneous memory systems in LLM inference.

\section{Background}

\begin{figure}[!t]
\centering
  \begin{subfigure}[b]{0.45\columnwidth}
    \includegraphics[width=\linewidth]{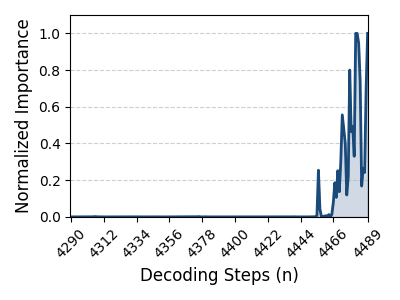}
    \caption{Token 2777 at layer 8.}
    \label{tok2777}
  \end{subfigure}%
\hfill
  \begin{subfigure}[b]{0.45\columnwidth}
    \includegraphics[width=\linewidth]{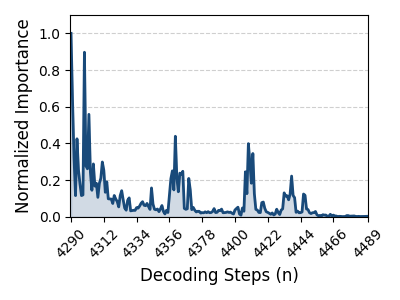}
    \caption{Token 4286 at layer 8.}
    \label{tok4286}
  \end{subfigure}%
\caption{The importance variation of 2 tokens during the decode stage.}
\label{fig:tokImpVar}
\end{figure}

\subsection{Token Importance Variation}
LLM inference comprises prefill and decode stages. In the prefill stage, the input prompt is processed in parallel to generate initial tokens, while the decode stage generates tokens auto-regressively, relying on the KV cache to store key and value vectors from prior tokens for efficient attention computation. This read-heavy behavior on KV cache in the decode stage places significant pressure on memory bandwidth and often becomes the dominant inference bottleneck~\cite{Agrawal2024SarathiServe}.

Recent methods mitigate full KV cache access via \textit{dynamic token bypassing}, selectively fetching influential tokens for attention. This leverages the unequal contributions of past tokens to the current one, reducing bandwidth usage while preserving accuracy. However, \textit{token importance varies over time}, with relevance shifting during the decode stage. To demonstrate, we analyzed attention scores from LLaMA-3.1-8B~\cite{MetaAI2024Llama318B} on LongBench~\cite{bai2024longbenchbilingualmultitaskbenchmark}, examining the attention heat-maps and selecting two representative tokens that show the attention variation during the inference at layer 8. Attention scores are used as proxies for importance: higher values indicate greater influence on the current token generation. As shown in Fig.~\ref{fig:tokImpVar}, token importance fluctuates substantially across decoding steps, with alternating peaks and troughs that reflect evolving contextual dependencies.


Due to this temporal variability, dynamic token bypassing strategies must ensure that all KV cache entries remain accessible in memory, even those being determined as not important at the moment, since their relevance may increase later. As a result, although bypassing reduces bandwidth pressure, it does not reduce memory capacity requirements. This limitation is becoming increasingly significant: as model sizes grow and techniques such as chain-of-thought prompting extend sequence lengths, the total KV cache footprint expands rapidly. Meanwhile, the capacity of high-bandwidth memory technologies such as HBM is improving at a much slower pace. This growing disparity underscores the need to reconsider how KV cache accessibility is managed in memory systems.

\subsection{Heterogeneous Memory Architecture for LLM}
To meet the growing memory demands of LLM inference amid limited HBM capacity, a natural architectural direction is to adopt heterogeneous memory systems that incorporate multiple memory technologies (e.g., HBM, LPDDR, DDR), each with distinct bandwidth and cost trade-offs. Recent advances in memory interface speeds and interconnect technologies have significantly increased the practical bandwidth of external memory. For example, LPDDR5X devices can reach speeds of up to 8,533 MT/s, and NVLink 4.0 supports up to 100~GB/s per link. These improvements have reduced the bandwidth gap between HBM and external memory to within an order of magnitude. A prominent example is NVIDIA’s GH200 Grace Hopper Superchip \cite{NVIDIAGraceHopper}, which combines a Hopper GPU with 96~GB of HBM3 delivering up to 4~TB/s of bandwidth, and a Grace CPU connected to 512~GB of LPDDR5X. The GPU and CPU are linked via NVLink-C2C, offering up to 900~GB/s of bandwidth. This design allows the GPU to directly access CPU-attached LPDDR5X memory at high speed, effectively extending both capacity and usable memory bandwidth beyond what HBM alone can provide.

As the bandwidth disparity between HBM and external memory reduces, it becomes increasingly beneficial to utilize the aggregate bandwidth of the entire heterogeneous memory system. Efficiently distributing inference-time datasets, such as KV cache entries, across these memory devices can significantly impact system performance, emphasizing the need for intelligent memory management strategies that consider both bandwidth and capacity constraints.

\section{Dynamic Data In-Memory Placement}
Despite the growing memory demands of LLM inference, the problem of dynamically placing inference-time datasets, particularly KV cache, across heterogeneous memory systems remains underexplored. This section introduces a formal framework for formulating the data placement problem in this context. Rather than proposing a specific scheduling algorithm, our objective is to characterize the design space by deriving a theoretical upper bound on speed performance. While not intended as a deployable policy, this upper bound quantifies the performance potential under idealized migration strategies, with the hope of inspiring future research on principled and efficient runtime data placement for LLM inference in heterogeneous memory systems.



\subsection{Mathematical Formulation of Inference Latency}\label{sec:formulation}
This subsection presents a mathematical formulation of LLM inference latency, which serves as the basis for evaluating the performance potential of KV cache placement strategies in heterogeneous memory systems. We focus exclusively on the \textit{decode stage}, where memory bandwidth is the primary bottleneck. This work centers on exploiting the variation of tokens' importance throughout the inference process. Specifically, model weights are accessed consistently across every inference step, rendering their placement in HBM essential for maximizing bandwidth utilization and minimizing latency overheads associated with these invariant accesses. Furthermore, in architectures employing techniques such as Mixture of Experts (MoE), KV cache assumes an even greater proportional role in the data access, amplifying its impact on performance and justifying targeted scheduling efforts. Hence, the full LLM model resides permanently in HBM and the only scheduling variable is the placement of KV cache entries across the memory hierarchy.

We model the decode stage as a sequence of inference steps, each indexed by a pair $(n, l)$, where $n$ denotes the token being generated and $l$ represents the transformer layer, which includes both Multi-Head Attention (MHA) and Multi-Layer Perceptron (MLP) submodules. Let $t_{(n,l)}$ represent the latency of step $(n,l)$. The total inference latency $T$ is the cumulative latency across all steps:
\begin{equation}
\label{eq:total_latency}
T = \sum_{\forall (n, l)} t_{(n,l)}.
\end{equation}

Each step may involve data movement between the processor and both HBM and off-package DRAM. Let $t^h_{(n,l)}$ and $t^e_{(n,l)}$ denote the latencies associated with accessing HBM and off-package DRAM, respectively. The step latency is defined as the maximum of the two:
\begin{equation}
\label{eq:step_latency}
t_{(n,l)} = \max\left\{ t^h_{(n,l)}, t^e_{(n,l)} \right\}.
\end{equation}

We now define $t^h_{(n,l)}$ and $t^e_{(n,l)}$ based on the KV cache placement and migration at each step. Let $\mathcal{S}_{(n,l)} = \{M^o_{(n,l)}, M^i_{(n,l)}, H^w_{(n,l)}, E^w_{(n,l)}\}$ represent the scheduling decision at step $(n,l)$:
\begin{itemize}
  \item $M^o_{(n,l)}$: KV cache data migrated from HBM to off-package DRAM,
  \item $M^i_{(n,l)}$: KV cache data migrated from off-package DRAM to HBM,
  \item $H^w_{(n,l)}$, $E^w_{(n,l)}$: newly written KV cache entries to HBM and off-package DRAM, respectively.
\end{itemize}
Let $H^r_{(n,l)}$ denote the data read from HBM for inference, and $B_h$ the HBM bandwidth. The latency of HBM accesses can be expressed as
\begin{equation}
\label{eq:hbm_latency}
t^h_{(n,l)} = \frac{ \left| H^r_{(n,l)} \right| + \left| H^w_{(n,l)} \right| + \left| M^i_{(n,l)} \right| + \left| M^o_{(n,l)} \right| }{ B_h },
\end{equation}
where $|\cdot|$ denotes the volume of data transferred. We assume that off-package DRAM is connected via full-duplex serial links (e.g., NVLink, UALink), with uni-directional bandwidth $B_k$, and internal DDR/LPDDR DRAM channels providing bandwidth $B_d$. Let $E^r_{(n,l)}$ be the data read from off-package DRAM for inference. The corresponding access latency is:

{\small
\begin{multline}
\label{eq:dram_latency}
t^e_{(n,l)} = \frac{ \left| E^r_{(n,l)} \right| }{ \min\{B_k, B_d\} } + 
\max\Biggl\{
\frac{ \left| E^w_{(n,l)} \right| + \left| M^o_{(n,l)} \right| }{ B_k },
\frac{ \left| M^i_{(n,l)} \right| }{ B_k },\\
\frac{ \left| E^w_{(n,l)} \right| + \left| M^i_{(n,l)} \right| + \left| M^o_{(n,l)} \right| }{ B_d }
\Biggr\}.
\end{multline}
}

Under each scheduling decision $\mathcal{S}$, the placement of KV cache data affects $H^r_{(n,l)}$ and $E^r_{(n,l)}$, and thereby alters the step latencies $t^h_{(n,l)}$ and $t^e_{(n,l)}$. As a result, the total inference latency $T$ varies with KV cache placement.

Let $P_H$ denote the capacity usage percentage of HBM, and assume off-package DRAM capacity is sufficiently large to hold all the KV cache. The goal of placement scheduling is to minimize the total inference latency while respecting HBM capacity limits. The optimization problem is thus formulated as:
\begin{equation}
\label{eq:opt_problem}
\begin{aligned}
\min_{\mathcal{S}} \quad & \sum_{\forall (n, l)} \max\left\{ t^h_{(n, l)}, t^e_{(n, l)} \right\} \\
\text{s.t.} \quad & P_H \le 100\%
\end{aligned}
\end{equation}

\subsection{Theoretical Upper Bound Exploration}\label{sec:upperbound}
The optimization problem, as defined in Section~\ref{sec:formulation}, encompasses a vast solution space, making exhaustive enumeration of all scheduling decisions at each step infeasible. Therefore, we search for a near-optimal solution to quantify the theoretical upper bound by adopting the Simulated Annealing~(SA)~\cite{Kirkpatrick1983Optimization} algorithm to explore the solution space. Since we are deducing the theoretical upper bound, we assume that the attention access patterns at each step in the decode stage are {\it a priori} known. However, it is impossible to scan the entire future steps' attention access patterns and decide the placement plan at each step. Two tunable parameters are exposed to the SA optimizer: $W$, the number of future decoding tokens to evaluate, which controls the exploration of the vast search space; and $R\in[0,1]$, the ratio of actual KV caches migration in all KV caches that can be migrated based on evaluation information to control the migration overhead. 

At each step, the SA algorithm collects the KV caches that will be accessed in the next $W$ tokens, ranks them based on their access frequency in a priority queue, and weighs the corresponding migration overhead by only selecting the top-$R$ portion of all KV caches qualified for migration. Let $T(W,R)$ denotes the objective (total decode latency); then $\Delta T=T_{new}-T_{current}$ is the cost increase of the proposed neighbour, and $C$ is the current “temperature” that monotonically decreases during cooling. The Metropolis rule $P(accept) = exp[\frac{-\Delta T}{C}]$ therefore gives the probability of accepting a worse move, which falls as either $\Delta T$ grows or $C$ cools.

In each annealing iteration the optimizer samples one of three proposal operators with probabilities $(0.4, 0.4, 0.2)$, respectively: (i) a \emph{window move} $\Delta W$ that perturbs the look-ahead window by $\{ \pm 1,\pm 2 \}$ while keeping $R$ fixed, (ii) a \emph{ratio move} $\Delta R$ that perturbs the migration ratio by $\{ \pm0.1 \}$ while keeping $W$ fixed, and (iii) a \emph{diagonal move} $\Delta W \Delta R$ that applies one perturbation of each kind simultaneously. Because $80\%$ of proposals touch exactly one parameter, the algorithm provides a clear attribution of each accepted improvement to either $W$ or $R$; the remaining $20\%$ of diagonal moves enable the algorithm to cross ridges where only coordinated adjustments of $W$ and $R$ reduce latency. For the cooling mechanism, we initialize the target initial acceptance ratio~$(p_0=0.8)$ high enough that uphill moves are accepted. The temperature is then reduced with the cooling rate~$(\alpha=0.9)$, so the acceptance probability of worse solutions declines smoothly and the walk shifts from exploration to exploitation. The process terminates when the best latency improves by less than a predefined threshold~($0.1\%$) across successive temperature levels, when the temperature falls below a minimal cutoff, or when a fixed iteration budget is reached, ensuring a balanced trade-off between solution quality and runtime.

Based on the assumption of known attention access patterns, the algorithm can place every KV cache entry in the optimal location before the compute phase demands it. The performance it reports constitutes a theoretical upper bound that no real-time policy, lacking such foresight, can outperform it. This upper bound analysis quantifies the headroom left by simpler heuristics.

\section{Evaluation}

\subsection{Methodology}
To evaluate the potential of optimizing KV cache placement for accelerating LLM inference, we implemented a behavioral simulator that models memory accesses during the decode stage of inference under a heterogeneous memory hierarchy. We assume that inference performance is primarily constrained by memory bandwidth, enabling us to estimate total inference latency using the formulation provided in Section~\ref{sec:formulation}. The memory system configuration is based on the NVIDIA GH200 Grace Hopper Superchip~\cite{NVIDIAGraceHopper}, as detailed in Table~\ref{tab:memory-module-config}.

\begin{table}[hbtp]
  \centering
  \caption{Memory system configuration.}
  \label{tab:memory-module-config}
  \renewcommand{\arraystretch}{1.2}
  \begin{tabular}{@{} p{0.3\linewidth} p{0.3\linewidth} r @{}}  
    \toprule
    \textbf{Component} & \textbf{Configuration} & \textbf{Value} \\ 
    \midrule
    HBM & Bandwidth & 4.9 TB/s \\
        & Capacity  & 24 GB    \\
    \addlinespace[0.2em]
    \midrule
    \addlinespace[0.2em]
    Off-package DRAM & Link bandwidth & 900 GB/s \\
                     & DRAM bandwidth & 500 GB/s \\
                     & Capacity       & 480 GB   \\
    \bottomrule
  \end{tabular}
\end{table}

We base our experiments on the LLaMA-3.1-8B model~\cite{MetaAI2024Llama318B}, which has a model size of approximately 16~GB, leaving around 8~GB of HBM capacity available for KV cache. To explore scenarios where KV cache capacity requirements exceed the available HBM space, we construct a long-context inference workload. Specifically, we feed the LLaMA-3.1-8B model with prompts of approximately 30k tokens from the NarrativeQA dataset of the LongBench benchmark~\cite{bai2024longbenchbilingualmultitaskbenchmark} and allow it to autoregressively decode 10K tokens. We record the layer-wise attention scores generated during the decode stage to obtain realistic access patterns for our memory-system simulations.


For comparison, we evaluate the following three KV cache placement strategies:

\begin{itemize}
  \item \textbf{Unlimited HBM}: Assumes unlimited HBM capacity, storing all data including model weights and KV cache in HBM for maximum bandwidth access. This idealized scenario yields optimal performance but is unrealistic.
  
  \item \textbf{Static Placement}: KV cache entries are written once without subsequent migration. New entries fill HBM until capacity is reached, after which they are placed in off-package DRAM, with no dynamic relocation.

  \item \textbf{Reactive Scheduling}: Upon accessing a KV cache entry absent from HBM, it is promoted to HBM. If HBM is full, the least recently used (LRU) entry is evicted to off-package DRAM, adapting placement to observed reuse patterns.

  \item \textbf{Page Granularity Scheduling}: Emulates the Quest approach~\cite{pmlr-v235-tang24l} by managing KV cache at page granularity (page size: 16). Entire pages are migrated with perfect foresight of token importance, though this incurs overhead from including unimportant tokens in the same page.

  \item \textbf{SA-Guided Scheduling}: Uses simulated annealing (SA), as described in Section~\ref{sec:upperbound}, with a priori access statistics to optimize KV cache placement. This upper-bound strategy leverages perfect foresight of token importance, highlighting potential gains over baselines via advanced scheduling.

\end{itemize}

\begin{figure}[!t]
  \centering
  \includegraphics[
    width=\columnwidth, scale=0.8
  ]{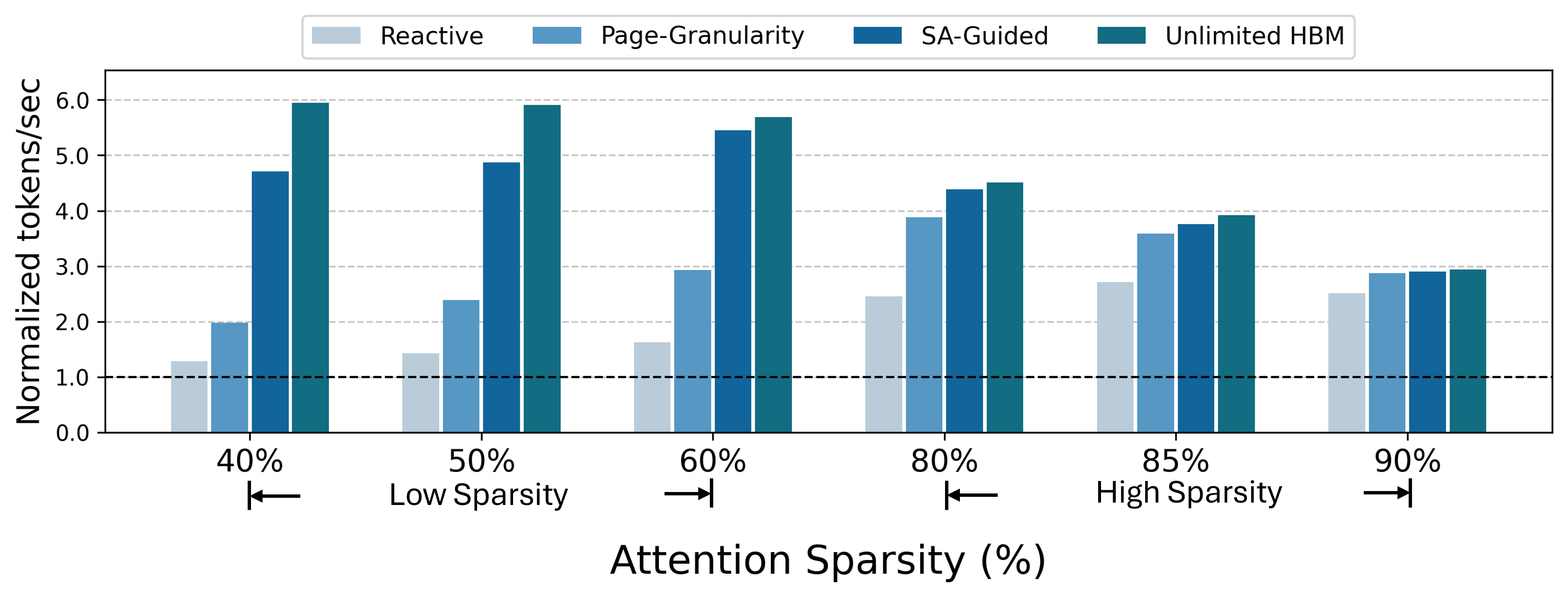}
  \captionsetup{justification=centering}
  \caption{The normalized tokens per second of strategies compared to the Static Placement.}
  \label{throughput}
\end{figure}


\subsection{Sensitivity Study}
To assess the impact of key factors on KV cache placement strategies, we examine sensitivity to attention sparsity and token importance variation. Attention sparsity refers to the fraction of past tokens excluded from attention computation at each decoding step; higher sparsity reduces computation and memory access overhead. Token importance variation measures shifts in influential tokens across steps: low variation occurs when important tokens remain largely consistent between consecutive decodings, while high variation indicates substantial changes.

Fig.~\ref{throughput} presents normalized tokens per second for the five strategies, relative to \textit{Static Placement}, across varying attention sparsity levels. At low sparsity (e.g., below 60\%), \textit{Reactive Scheduling} and \textit{Page Granularity Scheduling} yield only marginal gains. This stems from excessive migration overhead: dense token importance inflates the working set beyond HBM capacity, triggering frequent migrations that compete with inference accesses and reduce bandwidth efficiency. In contrast, high sparsity (e.g., above 80\%) shrinks the active KV cache footprint, minimizing migration costs and narrowing the performance gap between \textit{SA-Guided Scheduling} and other strategies.

\begin{figure}[!t]
  \centering

  \begin{minipage}[t]{0.485\columnwidth}
    \centering
    \includegraphics[width=\linewidth]{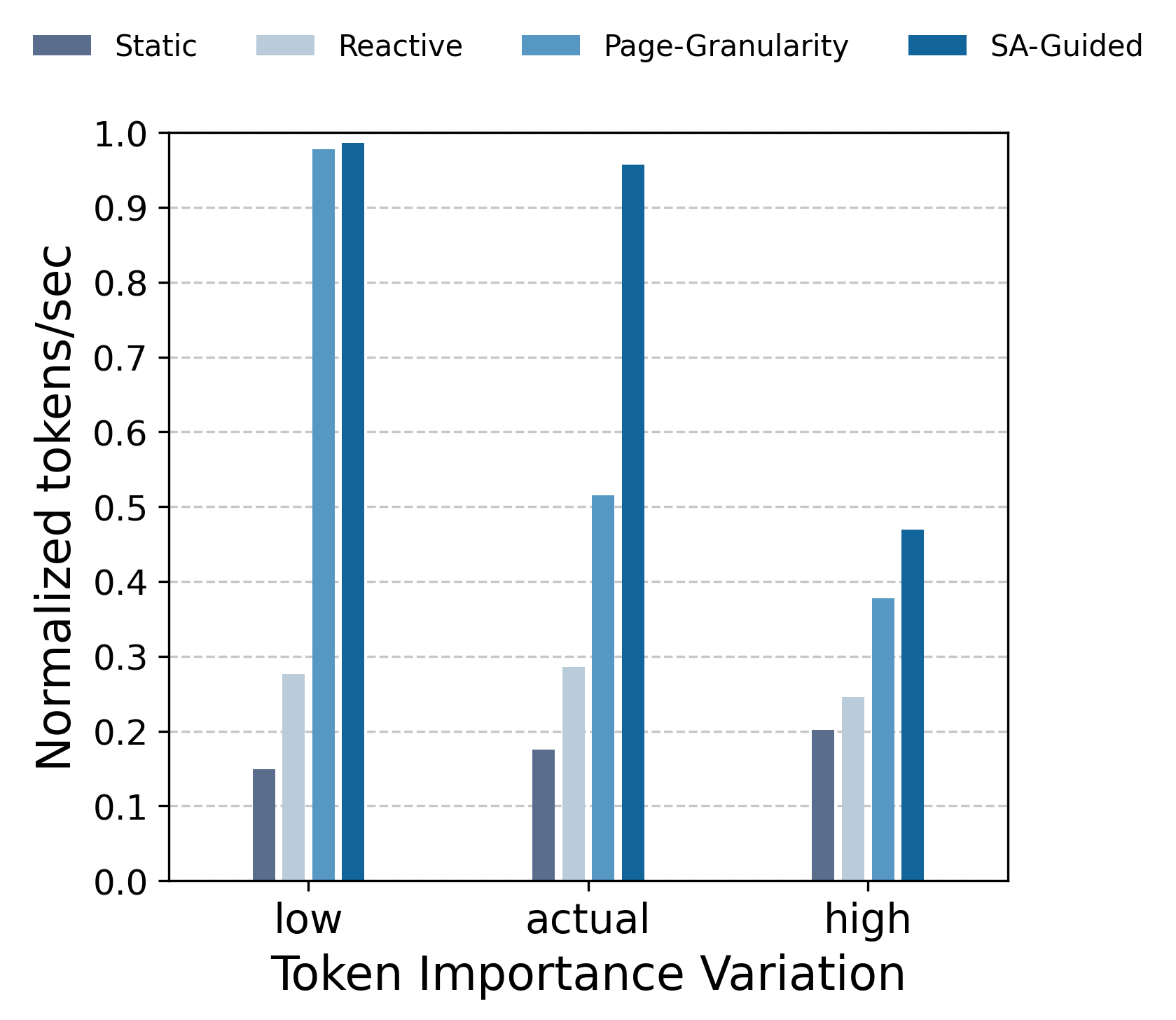}
    \captionsetup{justification=centering}
    \captionof{figure}{Normalized tokens/s vs. Unlimited HBM.}
    \label{fig:random}
  \end{minipage}\hfill
  \begin{minipage}[t]{0.485\columnwidth}
    \centering
    \includegraphics[width=\linewidth]{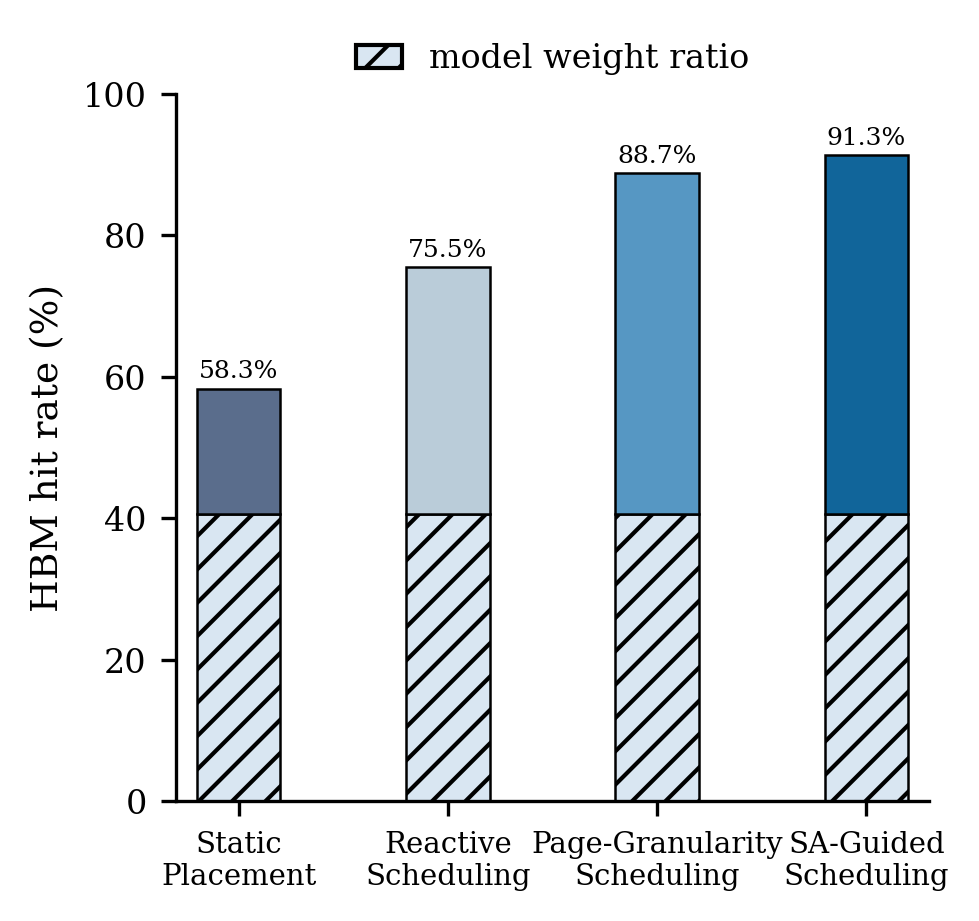}
    \captionsetup{justification=centering}
    \captionof{figure}{The HBM hit rates (attention sparsity = 60\%).}
    \label{fig:hit-rate}
  \end{minipage}

\end{figure}

To evaluate token importance variation, we synthesized two trace files simulating low and high variation scenarios during LLM inference. Fig.~\ref{fig:random} shows normalized tokens per second relative to \textit{Unlimited HBM} at 60\% attention sparsity. Under low variation, adaptive strategies like \textit{Page Granularity Scheduling} and \textit{SA-Guided Scheduling} approach optimal performance, as stable token importance requires minimal migrations for effective bandwidth utilization. Conversely, high variation degrades these strategies due to frequent migrations needed to track shifting access patterns. Overall, dynamic in-memory placement offers greater benefits for inference tasks with low importance variation.

Across all scenarios, \textit{SA-Guided Scheduling} consistently outperforms (4$\times$ to 5$\times$) the baseline by leveraging foresight to minimize migrations and optimize aggregate bandwidth. The HBM hit rate is also controlled at an efficient ratio as shown in Fig~\ref{fig:hit-rate}. The wide gap to baseline strategies highlights substantial potential for advanced techniques (e.g. predictive modeling, reinforcement learning, or online learning of token access patterns) to bridge this divide.

\section{Conclusion}
This work presents the first formal exploration of dynamic KV cache placement for LLM inference in heterogeneous memory systems. Instead of proposing a specific scheduling policy, we mathematically formulate the placement problem and derive a theoretical upper bound using simulated annealing, assuming perfect knowledge of token importance. Simulation results reveal up to a 5.87$\times$ performance gap between this upper bound and simple baseline strategies, underscoring the significant headroom for improvement. By making this gap explicit, our study provides a foundation and motivation for future research on practical, adaptive scheduling techniques that can approach this upper bound and better harness the capabilities of heterogeneous memory architectures.

\bibliographystyle{IEEEtran}
\bibliography{reference} 

\vfill

\end{document}